\documentclass[aps,prl,showpacs,twocolumn]{revtex4}
\usepackage{amsmath, amssymb}
\usepackage{graphicx}
\usepackage{subfigure}

\DeclareMathOperator{\tr}{Tr}
\DeclareMathOperator{\re}{Re}

\renewcommand{\vec}[1]{\mathbf{#1}}

\begin{document}

\title{Inseparability criteria for bipartite quantum states}

\author{E. Shchukin}
\email{evgeny.shchukin@uni-rostock.de}
\author{W. Vogel}
\email{werner.vogel@uni-rostock.de} \affiliation{Arbeitsgruppe Quantenoptik, Institut f\"ur Physik,
Universit\"at Rostock, D-18051 Rostock, Germany}

\begin{abstract}
We provide necessary and sufficient conditions for the partial
transposition of bipartite harmonic quantum states to be
nonnegative. The conditions are formulated as an infinite series of
inequalities for the moments of the state under study. The violation
of any inequality of this series is a sufficient condition for
entanglement. Previously known entanglement conditions are shown to
be special cases of our approach.
\end{abstract}


\maketitle

Entanglement plays a key role in the rapidly developing field of
quantum information processing. In this context it is important to
provide methods for characterizing entangled quantum states on the
basis of observable quantities. However, already in seemingly simple
cases this problem turns out to be rather complex.  Even for a
two-party harmonic-oscillator system so far there exists no complete
characterization of entanglement to be used in experiments.

For characterizing entanglement, that is inseparability, of the
density operator of a bipartite continuous variable system, one may use the
Peres-Horodecki condition~\cite{prl-77-1413, pla-223-1, pla-232-333}. A
sufficient condition for entanglement consists in the negativity of the
partial transposition (NPT) of the quantum state of the two-party system. To
characterize NPT for such a system completely, however, to our best knowledge
is still an unsolved problem.

A sufficient condition for the NPT has been proposed by
Simon~\cite{prl-84-2726}. It is based on second-order moments of
position and momentum operators. For the special case of Gaussian
states the resulting entanglement criterion has been shown to be
necessary and sufficient. Another inseparability condition based on
second moments has been derived without explicitely using the NPT
condition~\cite{prl-84-2722}. This condition is also complete for
the characterization of entanglement of two-mode Gaussian states.
The latter approach has been extended to special higher-order
moments~\cite{prl-88-120401, quant-ph-0501012} and even to more
general operator functions~\cite{pra-67-052104}.

The complexity of the problem under study may become clear when we go back to
a related but simpler problem. It consists in the characterization of
nonclassical effects based on the negativity of the Glauber-Sudarshan
$P$-function. Only recently the problem was solved of how to characterize the
nonclassicality in terms of observable quantities. This requires an infinite
hierarchy of conditions formulated either in terms of characteristic
functions~\cite{prl-89-283601} or in terms of observable
moments~\cite{pra-71-011802, quant-ph-0506029}.

In the present contribution we will further develop the concept of
the complete characterization of single-mode nonclassicality with
the aim to characterize the entanglement of bipartite continuous
variable quantum states. Based on the NPT condition we derive a
hierarchy of necessary and sufficient conditions for the NPT in
terms of observable moments. Even though this only leads to
sufficient conditions for entanglement, it can be applied to a
variety of quantum states.  It does not only contain as a special
case the condition of Simon~\cite{prl-84-2726}, but also other types
of inequalities~\cite{prl-84-2722, prl-88-120401, quant-ph-0501012,
pra-67-052104}. For the sake of strictness of the conclusions we
restrict our attention to a two-mode harmonic oscillator. Hence the
derived conditions are useful for radiation modes, motional states
of trapped ions and related systems. The further extension of our
approach to more general continuous variable systems seems feasible,
but this requires some further investigations.

A bipartite quantum state $\hat{\varrho}$ is called separable if it
is a convex combination of factorizable states,
\begin{equation}\label{eq:rhosep}
    \hat{\varrho} = \sum^{+\infty}_{n=0} p_n \hat{\varrho}^{(n)}_1 \otimes \hat{\varrho}^{(n)}_2,
\end{equation}
where $\hat{\varrho}^{(n)}_1$ and $\hat{\varrho}^{(n)}_2$ are states
of the first and the second part respectively. The $p_n$ are
nonnegative numbers satisfying $\sum^{+\infty}_{n=0} p_n = 1$ and 
the series in \eqref{eq:rhosep} is convergent in the trace norm. A state which cannot be
represented in the form \eqref{eq:rhosep} is called entangled.

It is well known that the full transposition of any (one-partite or
multi-partite) state is again a quantum state. But in the
multipartite case, along with the notion of the full transposition,
there is a notion of partial transposition, which transposes only
some parts of the state and leaves the others unchanged. In contrast
to the full transposition, partial transposition does not
necessarily map a quantum state onto a quantum state. Clearly, the
partially transposed density operator $\hat{\varrho}^{\mathrm{PT}}$
of a separable state $\hat{\varrho}$, Eq.~\eqref{eq:rhosep}, is
again a quantum state (we consider transposition of the second part;
our results are unchanged if we transpose the first part):
\begin{equation}\label{eq:rhopt}
    \hat{\varrho}^{\mathrm{PT}} = \sum^{+\infty}_{n=0} p_n \hat{\varrho}^{(n)}_1 \otimes
    \hat{\varrho}^{(n)\mathrm{T}}_2,
\end{equation}
since $\hat{\varrho}^{(n)\mathrm{T}}_2$ is a quantum state for all
$n$ and so is $\hat{\varrho}^{\mathrm{PT}}$ according to the
expansion \eqref{eq:rhopt}. This property of separable states allows
us to use the NPT as a sufficient condition for inseparability. This
condition is referred to as Peres-Horodecki condition
\cite{prl-77-1413, pla-223-1, pla-232-333}. In general the partial
transposition of a density operator may no longer be nonnegative
defined. It is this property that we will use for the derivation of
our hierarchy of conditions.

Remember that a Hermitian operator $\hat{A}$ is called nonnegative if
\begin{equation}
    \langle \psi |\hat{A}| \psi \rangle \geqslant 0
\end{equation}
holds for all states $|\psi\rangle$.  The left hand side of this inequality
can be written as
\begin{equation}\label{eq:ff}
\langle \psi |\hat{A}| \psi \rangle =
    \tr\Bigl(\hat{A}|\psi\rangle\langle\psi|\Bigr).
\end{equation}
The nonnegative operator $|\psi\rangle\langle\psi|$ can be
represented in the form $\hat{f}^\dagger \hat{f}$, where $\hat{f}$
may be written as
\begin{equation}\label{eq:fvac}
    \hat{f} = |\mathrm{vac}\rangle \langle \psi|.
\end{equation}
The state $|\mathrm{vac}\rangle$ is the vacuum state, in our
bipartite case it reads, in the Fock basis, as $|\mathrm{vac}\rangle
= |00\rangle$. Any pure bipartite state $|\psi\rangle$ can be
expressed as
\begin{equation}
    |\psi\rangle = \hat{g}^\dagger |00\rangle,
\end{equation}
for an appropriate operator function $\hat{g} = g(\hat{a},
\hat{b})$, where $\hat{a}$ and $\hat{b}$ are the annihilation
operators of the first and the second mode respectively. Hence the
operator~\eqref{eq:fvac} is of the form
\begin{equation}\label{eq:fvac2}
    \hat{f} = |00\rangle \langle 00| \hat{g}.
\end{equation}
The two-mode vacuum density operator may be expressed in the normally-ordered form,
\begin{equation}\label{eq:twomode-vac}
    |00\rangle \langle 00|\equiv \;\; : \exp(-\hat{a}^{\dagger} \hat{a} -
     \hat{b}^{\dagger}  \hat{b}) :\,,
\end{equation}
with $:\dots :$ denoting the normally-ordering prescription. By
combining Eqs~\eqref{eq:fvac2} and~\eqref{eq:twomode-vac} it
immediately follows that the normally-ordered form of the resulting
operator $\hat{f}$ exists. Hence we conclude: \textit{a Hermitian
operator is nonnegative if and only if for any operator $\hat{f}$
whose normally-ordered form exists the inequality}
\begin{equation}\label{eq:ff2}
    \langle \hat{f}^\dagger \hat{f} \rangle_{\hat{A}} =
    \tr\bigl(\hat{A} \hat{f}^\dagger \hat{f}\bigr) \geqslant 0
\end{equation}
\textit{is satisfied}.

Now we can use the inequality \eqref{eq:ff2} to apply the
Peres-Horodecki condition, which gives the following result.
\textit{For any separable state $\hat{\varrho}$ the inequality}
\begin{equation}\label{eq:ff3}
    \langle \hat{f}^\dagger \hat{f} \rangle^{\mathrm{PT}} \equiv \langle
    \hat{f}^\dagger \hat{f} \rangle_{\hat{\varrho}^{\mathrm{PT}}} =
    \tr\bigl(\hat{\varrho}^{\mathrm{PT}} \hat{f}^\dagger \hat{f}\bigr) \geqslant 0
\end{equation}
\textit{holds true for any operator $\hat{f}$ whose normally-ordered form
exists}. According to the latter assumption the operator $\hat{f}$ can be expanded as
\begin{equation}\label{eq:f}
    \hat{f} = \sum^{+\infty}_{n, m, k, l = 0} c_{nmkl}
    \hat{a}^{\dagger n} \hat{a}^m \hat{b}^{\dagger k} \hat{b}^l.
\end{equation}
Upon substituting this expansion into the inequality~\eqref{eq:ff3} we get the condition
\begin{equation}\label{eq:ffPT}
    \langle \hat{f}^\dagger \hat{f} \rangle^{\mathrm{PT}} =
    \sum^{+\infty}_{\substack{n, m, k, l = 0\\ p, q, r, s = 0}} c^*_{pqrs} c_{nmkl}
    M_{pqrs,nmkl} \geqslant 0,
\end{equation}
for the moments of the partial transposition
\begin{equation}\label{eq:MPT}
    M_{pqrs, nmkl} = \langle \hat{a}^{\dagger q} \hat{a}^p \hat{a}^{\dagger n} \hat{a}^m
    \hat{b}^{\dagger s} \hat{b}^r \hat{b}^{\dagger k} \hat{b}^l \rangle^{\mathrm{PT}}.
\end{equation}
The left hand side of the inequality~\eqref{eq:ffPT} is a quadratic
form with respect to the coefficients $c_{nmkl}$ of the
expansion~\eqref{eq:f}. The Silvester criterion states that the
inequality \eqref{eq:ffPT} holds for all $c_{nmkl}$ if and only if
all the main minors of the from \eqref{eq:ffPT} are nonnegative. To
write these minors explicitly we need to find the relation between
the moments \eqref{eq:MPT} of the partially transposed state and
those of the original one.

It is convenient to number multi-indices $(n, m, k, l)$ by a single
number so that the moments $M_{pqrs, nmkl}$ be numbered by two
indices like matrix elements are. The problem of numbering the
multi-indices $(n, m, k, l)$ is equivalent to the ordering of the
set $\{(n, m, k, l)\}$ of all such multi-indices. Then we can use
the ordinal number of a multi-index $(n, m, k, l)$ in the ordered
sequence of multi-indices. We use the following order: for any two
multi-indices $\vec{u} = (n, m, k, l)$ and $\vec{v} = (p, q, r, s)$
\begin{equation}
    \vec{u} < \vec{v} \Leftrightarrow
    \begin{cases}
        |\vec{u}| < |\vec{v}|,\ \text{or} \\
    |\vec{u}| = |\vec{v}|\ \text{and}\ \vec{u} <^\prime \vec{v},
    \end{cases}
\end{equation}
where $|\vec{u}| = n+m+k+l$ and $\vec{u} <^\prime \vec{v}$ means that the
first nonzero difference $r-k, s-l, p-n, q-m$ is positive. The resulting
ordered sequence of the moments starts as follows:
\begin{equation}\label{eq:ordseq}
\begin{split}
    &1, \langle \hat{a} \rangle, \langle \hat{a}^\dagger \rangle, \langle \hat{b} \rangle, \langle \hat{b}^\dagger \rangle, \langle \hat{a}^2 \rangle, \langle \hat{a}^\dagger \hat{a} \rangle,
    \langle \hat{a}^{\dagger 2} \rangle, \langle \hat{a} \hat{b} \rangle, \langle \hat{a}^\dagger \hat{b} \rangle, \\
    &\langle \hat{b}^2 \rangle, \langle \hat{a} \hat{b}^\dagger \rangle,
    \langle \hat{a}^\dagger \hat{b}^\dagger \rangle, \langle \hat{b}^\dagger \hat{b} \rangle, \langle \hat{b}^{\dagger 2} \rangle, \ldots .
\end{split}
\end{equation}

One can easily check that the moments $M_{pqrs, nmkl}$, Eq.~\eqref{eq:MPT},
can be expressed in terms of the moments of the original state as follows
\begin{equation}
    \langle \hat{a}^{\dagger q} \hat{a}^p \hat{a}^{\dagger n} \hat{a}^m
    \hat{b}^{\dagger s} \hat{b}^r \hat{b}^{\dagger k} \hat{b}^l \rangle^{\mathrm{PT}} =
    \langle \hat{a}^{\dagger q} \hat{a}^p \hat{a}^{\dagger n} \hat{a}^m
    \hat{b}^{\dagger l} \hat{b}^k \hat{b}^{\dagger r} \hat{b}^s \rangle. \nonumber
\end{equation}
Note that these moments are linear combinations of normally-ordered ones,
$\langle \hat{a}^{\dagger n} \hat{a}^m \hat{b}^{\dagger k} \hat{b}^l \rangle$,
according to the following formula for the antinormally-ordered product
$\hat{a}^n \hat{a}^{\dagger m}$,
\begin{equation}
    \hat{a}^n \hat{a}^{\dagger m} = \sum^{\min(n, m)}_{k=0} \frac{n! m!}{k! (n-k)! (m-k)!}
    \hat{a}^{\dagger m-k} \hat{a}^{n-k}.
\end{equation}

The nonnegativity of all the minors of the quadratic form \eqref{eq:ffPT} can now
be written as
\begin{equation}\label{eq:DN}
    \forall N: \quad
    D_N =
    \begin{vmatrix}
        M_{11} & M_{12} & \ldots & M_{1N} \\
        M_{21} & M_{22} & \ldots & M_{2N} \\
        \hdotsfor{4} \\
        M_{N1} & M_{N2} & \ldots & M_{NN}
    \end{vmatrix}
    \geqslant 0,
\end{equation}
where the moments $M_{ij}$ defined in Eq.~\eqref{eq:MPT} are now given as
\begin{equation}\label{eq:MPT2}
    M_{ij} = \langle \hat{a}^{\dagger q} \hat{a}^p \hat{a}^{\dagger n} \hat{a}^m
    \hat{b}^{\dagger l} \hat{b}^k \hat{b}^{\dagger r} \hat{b}^s \rangle,
\end{equation}
with $(n, m, k, l)$ and $(p, q, r, s)$ being the $i$th
and the $j$th indices in the ordered sequence~\eqref{eq:ordseq}.
Now we can explicitly formulate a necessary and sufficient condition for the
positivity of partial transposition. \textit{The partial transposition of a
  bipartite quantum state is nonnegative, if and only if all the determinants}
\begin{equation}\label{eq:DN3}
    D_N =
    \begin{vmatrix}
        1 & \langle \hat{a} \rangle & \langle \hat{a}^\dagger \rangle &
        \langle \hat{b}^\dagger \rangle & \langle \hat{b} \rangle & \ldots \\
        \langle \hat{a}^\dagger \rangle & \langle \hat{a}^\dagger \hat{a} \rangle &
        \langle \hat{a}^{\dagger 2}\rangle &
        \langle \hat{a}^\dagger \hat{b}^\dagger \rangle & \langle \hat{a}^\dagger \hat{b} \rangle & \ldots \\
        \langle \hat{a} \rangle & \langle \hat{a}^2 \rangle & \langle \hat{a} \hat{a}^\dagger \rangle &
        \langle \hat{a} \hat{b}^\dagger \rangle & \langle \hat{a} \hat{b} \rangle & \ldots \\
        \langle \hat{b} \rangle & \langle \hat{a} \hat{b} \rangle & \langle \hat{a}^\dagger \hat{b}\rangle &
        \langle \hat{b}^\dagger \hat{b} \rangle & \langle \hat{b}^2 \rangle & \ldots \\
        \langle \hat{b}^\dagger \rangle & \langle \hat{a} \hat{b}^\dagger \rangle &
        \langle \hat{a}^\dagger \hat{b}^\dagger \rangle &
        \langle \hat{b}^{\dagger 2} \rangle & \langle \hat{b} \hat{b}^\dagger \rangle & \ldots \\
    \end{vmatrix}
\end{equation}
\textit{are nonnegative, that is}
\begin{equation}\label{eq:DN2}
    \forall N : \quad D_N \geqslant 0.
\end{equation}

\textit{The other way around, if there exists a negative determinant,}
\begin{equation}\label{eq:DN-neg}
    \exists N: \quad D_N < 0,
\end{equation}
\textit{the NPT has been demonstrated. Hence the condition~(\ref{eq:DN-neg})
  is a necessary and sufficient condition for the NPT. That is, it provides a
  sufficient condition for the state under consideration to be entangled}.
This result is the central finding of the present contribution.  It is
important to stress that all the moments occurring in our conditions can be
determined experimentally. This requires a straightforward extension of the
method, as proposed for measuring the moments of a single-mode radiation
field~\cite{quant-ph-0506029}, to the case of two modes.

Even though the above conditions~(\ref{eq:DN-neg}) are necessary and
sufficient for NPT, for practical applications it may be useful to
derive other types of conditions. They may be more efficient for
verifying entanglement of particular quantum states in practice,
violations of these new conditions may be found easier than the
negativities of the determinants~\eqref{eq:DN3}. For this purpose we
consider the following two possibilities:
\begin{enumerate}
\renewcommand{\theenumi}{\roman{enumi}}
\renewcommand{\labelenumi}{(\theenumi)}
\item \textit{Restricting the generality} of the operator \eqref{eq:f} by
  restricting the generality of the coefficients $c_{nmkl}$ leads to other
  conditions for inseparability. For example,
  setting some coefficients $c_{nmkl}$ to zero leads to conditions for the
  nonnegativity of subdeterminants of $D_N$, which are obtained by canceling
  rows and columns with the same indices.
\item \textit{Identifying some coefficients} $c_{nmkl}$ leads to another way
  to formulate new conditions for NPT by reducing the number of independent
  variables of the form~\eqref{eq:ffPT}. For example, the quadratic form
  derived for the operator $\hat{f} = c_1 + c_2 \hat{a} + c_3 \hat{b}$ depends
  on three independent (complex) variables $c_1$, $c_2$ and $c_3$.  If we
  identify the last two coefficients $c_2 = c_3$, i.e. $\hat{f} = c_1 +
  c_2(\hat{a}+\hat{b})$, we obtain a quadratic form of two variables $c_1$ and
  $c_2$. The resulting sufficient conditions for NPT can be formulated in
  terms of
  the nonnegativity of determinants whose entries are combinations of the
  moments~\eqref{eq:MPT2}.
\end{enumerate}
Examples for the usefulness of both types of new conditions will be considered
in the following.

Let us compare our approach with some previously known results. We start with
the separability condition found by Simon~\cite{prl-84-2726}. Define matrices $A_1$,
$A_2$, $C$ and $J$ by the expressions
\begin{equation}\label{eq:ABC}
\begin{split}
    A_i &=
    \begin{pmatrix}
        \langle (\Delta \hat{x}_i)^2 \rangle & \langle \{\Delta \hat{x}_i, \Delta \hat{p}_i\} \rangle \\
        \langle \{\Delta \hat{x}_i, \Delta \hat{p}_i\} \rangle & \langle (\Delta \hat{p}_i)^2 \rangle
    \end{pmatrix}, \\
    C &=
    \begin{pmatrix}
        \langle \Delta \hat{x}_1 \Delta \hat{x}_2 \rangle & \langle \Delta \hat{x}_1 \Delta \hat{p}_2 \rangle \\
        \langle \Delta \hat{p}_1 \Delta \hat{x}_2 \rangle & \langle \Delta \hat{p}_1 \Delta \hat{p}_2 \rangle
    \end{pmatrix},
\end{split}
\end{equation}
and $J = \left(\begin{smallmatrix} 0 & 1 \\ -1 & 0
  \end{smallmatrix}\right)$.
The Simon condition can be formulated as follows: for any separable
quantum state the quantity~$S$,
\begin{equation}\label{eq:Simon}
\begin{split}
    &S = \det A_1 \det A_2 + \left(\frac{1}{4} + \det C\right)^2 - \\
    &\tr\Bigl(A_1 J C J A_2 J C^T J\Bigr) -
    \frac{1}{4}\Bigl(\det A_1 + \det A_2\Bigr) \geqslant 0,
\end{split}
\end{equation}
is nonnegative. Since the position and momentum operators
$\hat{x}_i$ and $\hat{p}_i$, $i = 1, 2$, are linear combinations of
the creation and annihilation operators, the
condition~\eqref{eq:Simon} can also be formulated in terms of the
moments $M_{ij}$, Eq.~\eqref{eq:MPT2}. By using our concept the left
hand side of the inequality~\eqref{eq:Simon} reads much simpler in
terms of the moments $M_{ij}$. Simon's quantity $S$ is nothing but
our fifth order determinant $D_5$:
\begin{equation}
    S \equiv D_5.
\end{equation}
Therefore the Simon condition~\eqref{eq:Simon} is exactly one
condition of our hierarchy~\eqref{eq:DN2}.

Another separability condition has been formulated
in~\cite{prl-84-2722}. For any non-zero real parameter $r$ one may
define operators $\hat{u}$ and $\hat{v}$ via
\begin{equation}\label{eq:uv}
    \hat{u} = |r| \hat{x}_1 + r^{-1} \hat{x}_2, \quad
    \hat{v} = |r| \hat{p}_1 - r^{-1} \hat{p}_2.
\end{equation}
Then, for any separable state the inequality
\begin{equation}\label{eq:Duan}
    \langle (\Delta \hat{u})^2 \rangle + \langle (\Delta \hat{v})^2 \rangle- (r^2 + r^{-2})
    \geqslant 0
\end{equation}
holds true for all $r \not=0$. The condition~\eqref{eq:Duan} can be simplified
as follows. We express the operators $\hat{u}$ and $\hat{v}$ as linear
combinations of the creation and annihilation operators and substitute them
into the inequality~\eqref{eq:Duan}. Now its left hand side can be minimized
with respect to $r$. Then the inequality \eqref{eq:Duan} is found to be
equivalent to
\begin{equation}\label{eq:Duan2}
    \langle \Delta \hat{a}^\dagger \Delta \hat{a} \rangle
    \langle \Delta \hat{b}^\dagger \Delta \hat{b} \rangle \geqslant
    \re^2\langle \Delta \hat{a} \Delta \hat{b} \rangle.
\end{equation}
The inequality \eqref{eq:Duan2} is related to a subdeterminant of
$D_N$, to the one resulting from the operator $\hat{f} = c_1 + c_2
\hat{a} + c_3 \hat{b}$:
\begin{equation}\label{eq:d}
    d =
    \begin{vmatrix}
        1 & \langle \hat{a} \rangle & \langle \hat{b}^\dagger \rangle \\
        \langle \hat{a}^\dagger \rangle & \langle \hat{a}^\dagger \hat{a} \rangle &
        \langle \hat{a}^\dagger \hat{b}^\dagger \rangle \\
        \langle \hat{b} \rangle & \langle \hat{a} \hat{b} \rangle &
        \langle \hat{b}^\dagger \hat{b} \rangle
    \end{vmatrix}.
\end{equation}
The condition $d \geqslant 0$ explicitly reads as
\begin{equation}
    \langle \Delta \hat{a}^\dagger \Delta \hat{a} \rangle
    \langle \Delta \hat{b}^\dagger \Delta \hat{b} \rangle \geqslant
    |\langle \Delta \hat{a} \Delta \hat{b} \rangle|^2,
\end{equation}
from which it becomes clear that the condition~\eqref{eq:Duan} is a weaker
form of the nonnegativity of our determinant~\eqref{eq:d}.

There have been various other results in the literature. We can only briefly
comment on some of them. A generalization of the inequality~\eqref{eq:Duan} was
obtained in~\cite{pra-67-052104}.
The main result is not explicitly based on the nonnegativity of partial
transposition, but in fact it is a consequence of it. Using our theory we can
show that the main inequality follows from Eq.~\eqref{eq:ff3} for the operator
$\hat{f} = c_1 + c_2 \hat{A}_1 + c_3 \hat{B}_1 + c_4 \hat{A}_2 + c_5
\hat{B}_2$ (in the notation of work \cite{pra-67-052104}), which is readily
obtained by our approach~(ii). Another inequality has been obtained
in~\cite{prl-88-120401}.  It differs from the condition~\eqref{eq:Duan} and
its generalization~\cite{pra-67-052104} in that it gives a lower bound for the
product of the variances of the operators~\eqref{eq:uv} rather than for their
sum. Though this inequality does not explicitly use NPT, in fact it is also a
consequence of it.  The condition provided in~\cite{prl-88-120401} easily
follows from the nonnegativity of $d$, Eq.~\eqref{eq:d}, together with the determinant
corresponding to the operator $\hat{f} = c_1 + c_2(\hat{a}+\hat{b}) +
c_3(\hat{a}^\dagger + \hat{b}^\dagger)$. Yet another condition has been
recently obtained in~\cite{quant-ph-0501012}. It is easy to verify that this
condition is a weaker form of the nonnegativity of the determinant
corresponding to the operator $\hat{f} = c_1 + c_2 \hat{a} \hat{b} + c_3
\hat{a}^\dagger \hat{b}^\dagger$.

Let us illustrate our theory for the example of an entangled quantum
composed of two coherent states,
\begin{equation}\label{eq:psi3}
    |\psi_-\rangle =\mathcal{N}_- (\alpha, \beta)\Bigl(|\alpha, \beta\rangle - |-\alpha, -\beta\rangle\Bigr),
\end{equation}
where the normalization is
\begin{equation}
    \mathcal{N}_-(\alpha, \beta) = \Bigl(2(1-e^{-2(|\alpha|^2+|\beta|^2)})\Bigr)^{-1/2}.
\end{equation}
The violations of the conditions~\eqref{eq:Simon} and~\eqref{eq:Duan} fail to
demonstrate the entanglement of this state. However, we may apply the
following simple subdeterminant:
\begin{equation}
    s =
    \begin{vmatrix}
        1 & \langle \hat{b}^\dagger \rangle & \langle \hat{a} \hat{b}^{\dagger} \rangle \\
    \langle \hat{b} \rangle & \langle \hat{b}^\dagger\hat{b} \rangle & \langle \hat{a} \hat{b}^\dagger\hat{b} \rangle \\
    \langle \hat{a}^\dagger \hat{b} \rangle & \langle \hat{a}^\dagger \hat{b}^\dagger \hat{b} \rangle & \langle \hat{a}^\dagger \hat{a} \hat{b}^\dagger \hat{b} \rangle
    \end{vmatrix}.
\end{equation}
Explicitly it reads as
\begin{equation}
    s = -|\alpha|^2 |\beta|^4 \frac{\coth(|\alpha|^2 + |\beta|^2)}{\sinh^2(|\alpha|^2 + |\beta|^2)},
\end{equation}
which is clearly negative for all non-zero parameters and hence the
entanglement is verified.

Finally, we will consider another special kind of separability condition
obtained by our approach~(i). Let us consider the simplest subdeterminants of
the $D_N$, namely those of second order. They correspond to the
condition~\eqref{eq:ff3} with $\hat{f}$ being a linear combination of two
terms,
\begin{equation}\label{eq:f2}
    \hat{f} = c_1 \hat{a}^{\dagger n} \hat{a}^m \hat{b}^{\dagger k} \hat{b}^l + c_2 \hat{a}^{\dagger p} \hat{a}^q \hat{b}^{\dagger r} \hat{b}^s.
\end{equation}
Explicitely the resulting separability condition reads as
\begin{equation}\label{eq:2}
\begin{split}
    &\langle \hat{a}^{\dagger m} \hat{a}^n \hat{a}^{\dagger n} \hat{a}^m \hat{b}^{\dagger l} \hat{b}^k \hat{b}^{\dagger k} \hat{b}^l \rangle
    \langle \hat{a}^{\dagger q} \hat{a}^p \hat{a}^{\dagger p} \hat{a}^q \hat{b}^{\dagger s} \hat{b}^r \hat{b}^{\dagger r} \hat{b}^s \rangle
    \geqslant \\
    &\Bigl|\langle \hat{a}^{\dagger m} \hat{a}^n \hat{a}^{\dagger p} \hat{a}^q \hat{b}^{\dagger s} \hat{b}^r \hat{b}^{\dagger k} \hat{b}^l \rangle\Bigr|^2.
\end{split}
\end{equation}
This leads to rather simple conditions including moments of higher
orders. Note that the conditions \eqref{eq:2} generalize those
derived recently in~\cite{hillery-zubairy}.

In conclusion, we have derived a hierarchy of conditions that is
necessary and sufficient for the negativity of the partial
transposition of the density operator of a bipartite harmonic
oscillator. In fact this provides a sufficient condition for
entanglement in terms of observable quantities, that can be used for
a variety of practical applications. We have also demonstrated that
several previously known conditions appear to be special cases of
our approach.

\end{document}